\documentclass[a4paper,11pt]{article}
\pdfoutput=1
\usepackage{pos}

\newcommand{\be}{\begin{equation}}
\newcommand{\ee}{\end{equation}}
\newcommand{\bea}{\begin{eqnarray}}
\newcommand{\eea}{\end{eqnarray}}

\title{Quark-lepton correlations in gauge anomaly free abelian extension of the Standard Model }
\author[a]{Pietro Colangelo}
\author*[a]{Fulvia De Fazio}
\author[a,b]{Davide Milillo}

\affiliation[a]{Istituto Nazionale di Fisica Nucleare, Sezione di Bari,  Via Orabona 4, 70125 Bari, Italy}
\affiliation[b]{Physics Department of Bari University and Polytechnic,  Via Orabona 4, 70125 Bari, Italy}

\emailAdd{pietro.colangelo@ba.infn.it}
\emailAdd{fulvia.defazio@ba.infn.it}
\emailAdd{davide.milillo@ba.infn.it}

\abstract{We study $b \to s \ell_1^+ \ell_2^-$ transitions, both  for the lepton flavour conserving  $\ell_1=\ell_2$ and violating case $\ell_1 \neq \ell_2$, in a minimal extension of the Standard Model  proposed in \cite{Aebischer:2019blw}.  In this framework,  the Standard Model (SM)  gauge group is enlarged by a new $U(1)^\prime$ component. The fermion  $U(1)^\prime$ charges are assigned in a generation-dependent way, and involve three rational parameters $\epsilon_{1,2,3}$ summing to zero  by the condition of  cancellation of the  gauge anomalies. Each  $\epsilon_i$ is  common to all fermions in a generation, which produces  correlations among quark and lepton observables.   The new neutral gauge boson $Z^\prime$ has flavour violating couplings to quarks and leptons.  For SM allowed processes,  small deviations with respect to the SM predictions are found: this is  a consequence of  a feature of the model where  quark and lepton sectors preclude each other large deviations from SM.  Lepton flavour violating processes are allowed at tree-level. The experimental upper bounds for the rates of the processes $\tau^- \to \mu^- \mu^+ \mu^-$, $\mu^- \to e^- \gamma$, $\mu^- \to e^- e^+ e^-$ and the $ \mu^- \to e^-$ conversion in nuclei  play a hierarchical role in constraining the branching fractions of lepton flavour violating $B$ and $B_s$ decays. }

\FullConference{The European Physical Society Conference on High Energy Physics (EPS-HEP2025)\\
7-11 July 2025\\
Marseille, France\\}

\begin{document}

\begin{flushright}
 %   {BARI-TH/25-779}
\end{flushright}
\maketitle

\section{Introduction}

The Standard Model (SM) is empirically successful in describing the particle interactions up to the  energies probed by the LHC. Nevertheless,  it is commonly believed to be an incomplete theory since it  leaves several  issues unsolved:   the number and  texture of the fundamental parameters, the fermion generations and mixing, it does not account for gravity. Physics beyond the SM (BSM) is sought for,   SM  being its low-energy limit.  
Hints of BSM phenomena  have emerged as tensions between SM predictions and measurements, most notably in the form of flavour anomalies. Although not yet conclusive, they can collectively suggest BSM effects. Flavour physics plays a crucial role in this search. Namely,  flavour-changing neutral current processes (FCNC) are very sensitive to new physics (NP)  contributions since they are forbidden at tree-level in SM   \cite{Buras:2020xsm}.  

The list of the  anomalies includes \cite{Capdevila:2023yhq,Colangelo:2024ped}:
\begin{itemize}
\item Discrepancies in Cabibbo-Kobayashi-Maskawa (CKM)  matrix elements,  with long-standing tensions between inclusive and exclusive determinations of  $|V_{cb}|$ and $|V_{ub}|$ \cite{HeavyFlavorAveragingGroupHFLAV:2024ctg} and a deficit in the  first row CKM  unitarity relation  \cite{Kitahara:2024azt};  
\item Deviations in FCNC $b \to s$ transitions: suppressed branching ratios of $B \to K \mu^+ \mu^-$ and $B_s \to \phi \mu^+ \mu^-$,   anomalies in $B \to K^* \mu^+ \mu^-$ angular distributions  \cite{HeavyFlavorAveragingGroupHFLAV:2024ctg};  
\item Lepton flavour universality violation in semileptonic $b \to c \ell \bar \nu_\ell$ decays, with the ratios $R(D^{(*)})$ exceeding the SM prediction by about $3\sigma$ \cite{HeavyFlavorAveragingGroupHFLAV:2024ctg}.
\end{itemize} 
As for the the muon anomalous magnetic moment $(g-2)_\mu$ \cite{Muong-2:2025xyk}, a crucial role in the SM prediction  is played  by the leading-order hadronic-vacuum-polarization contribution,  the  uncertainty  of which requires to be settled \cite{Aliberti:2025beg}.  
Though each  deviation  is not significant enough to unambiguously signal NP, all together they can motivate BSM scenarios. 
Most observables involve  SM parameters and nonperturbative QCD  quantities, and it is mandatory to reliably assess their associated uncertainties \cite{Colangelo:2024ped,Buras:2021nns,Buras:2022wpw}.

A minimal extension of SM with an additional $U(1)^\prime$ gauge symmetry and right-handed neutrinos (the ABCD model) has been proposed in  \cite{Aebischer:2019blw}. It assigns generation-dependent $U(1)^\prime$ charges to fermions, yielding correlations among quark and lepton observables while preventing large deviations from SM predictions. Within this framework, lepton flavour violating (LFV) processes, forbidden in SM,  represent an avenue to access  new physics. Here we summarize the main findings in \cite{Colangelo:2025nbd}, focusing on the  correlations between flavour-conserving (LFC) and LFV rare $B$ decays, as well as between $b \to s \ell_1^- \ell_2^+$ transitions and purely leptonic LFV transitions.\footnote{Recent studies of lepton flavour violating $b \to s$ modes are in \cite{Crivellin:2015mga,Crivellin:2015lwa,Crivellin:2015era,Cornella:2021sby,Bordone:2021usz,Panda:2024ygr,Becirevic:2024vwy}.}

\section{ ABCD model}\label{abcd-summary}
In the ABCD model \cite{Aebischer:2019blw}  the SM gauge group is enlarged with an extra $\text{U}(1)^\prime$ symmetry predicting a new neutral gauge boson $Z^\prime$ with coupling $g_{Z^\prime}$ and mass $M_{Z^\prime}$. Many NP models introduce a new $Z^\prime$ and differ for the  $\text{U}(1)^\prime$ charge ($z$-hypercharge) assignment \cite{Leike:1998wr,Langacker:2000ju,Appelquist:2002mw,Rizzo:2006nw}. In ABCD, fermions are assigned  peculiar  $z$-hypercharges chosen to cancel gauge anomalies through rational solutions of six anomaly cancellation equations. The charges are expressed as the SM hypercharge plus generation-dependent shifts $\epsilon_i$ constrained by the condition  $\epsilon_1+\epsilon_2+\epsilon_3=0$ to cancel the gauge anomalies.  

The model comprises heavy right-handed neutrinos (of Dirac type) and preserves the SM quark sector. After rotation from flavour to mass eigenstates, the $Z^\prime$ couplings become flavour non-universal and may generate flavour-changing neutral currents (FCNC) and lepton flavour violation (LFV) at tree-level. In particular,
\begin{itemize}
\item the $Z^\prime$ couplings to left-handed fermions depend on CKM  and Pontecorvo-Maki-Nakagawa-Sakata (PMNS)   matrices for quarks and leptons, respectively,   
\item the right-handed couplings involve additional mixing parameters $(\tilde s_{ij},\phi_{ij})$ in the quark sector and $(\tilde t_{ij},\phi_{ij})$ in the lepton sector,  
\item non-vanishing $\epsilon_{i}$ induce LFV couplings, namely in $\mu \to e$, $\tau \to e$, and $\tau \to \mu$ transitions.  
\end{itemize}
The new parameters of the model are
\be
g_{Z^\prime}, \quad M_{Z^\prime}, \quad \epsilon_1, \quad \epsilon_2 \, , \label{pars}
\ee
together with CKM and PMNS parameters and, depending on the scenario, the mixing angles and phases of right-handed fermions.  
Three benchmark scenarios have been selected \cite{Aebischer:2019blw}, with  parameters $\epsilon_{1,2}\,$ and the elements of the CKM/PMNS matrices:  the scenarios,  denoted as A, B and C, are respectively characterised by having 
\begin{enumerate}
\item[A:]  no  flavour violation for right-handed fermions  
\item[B:] flavour violation only for right-handed quarks   
\item[C:] flavour violation for both right-handed quarks and leptons.  
\end{enumerate}
We focus on the scenario A,  a minimal  set up  which already  predicts  correlations among quark and lepton observables,  and  among lepton flavour conserving and violating processes.

\section{Observables }\label{obs-summary}
In the ABCD model flavour changing neutral currents occur at tree-level via $Z^\prime$ exchange. Also lepton flavour violating  processes are allowed. We focus on $b \to s \ell_1^+ \ell_2^-$ transitions, both lepton flavour-conserving and violating, in particular on the  exclusive modes $B_s \to \ell_1^+ \ell_2^-$, $B \to K^* \ell_1^+ \ell_2^-$ and $B_s \to \phi \ell_1^+ \ell_2^-$. 
In the model, the low-energy Hamiltonian for such modes comprises current-current, penguin, magnetic and semileptonic operators:
\be
H^{\rm eff}=-\frac{4G_F}{\sqrt{2}} V_{tb}V_{ts}^* \sum_i \left[ C_i O_i + C_i^\prime O_i^\prime \right], \label{hamil}
\ee
where $G_F$ is the Fermi constant and   $V_{ij}$ are
elements of the CKM  matrix (doubly Cabibbo suppressed terms proportional to $V_{ub} V_{us}^*$ are neglected) \cite{Altmannshofer:2008dz}.
Among the operators in \eqref{hamil}, $O_{7,9,10}$ are dominant in the SM. In the ABCD model the primed operators (chiral partners of the unprimed ones)  are generated, and the Wilson coefficients are modified in a flavour-dependent way.

Considering  the decay rate of  $B_s \to \ell_1^- \ell_2^+$,  that depends only on the decay constant $f_{B_s}$,  in SM with $\ell_1=\ell_2$ only the operator $O_{10}$ contributes. In the ABCD model also $O_{9,10}^{(\prime)}$ contribute and LFV transitions  are allowed.
For $B \to K^*\ell_1^+ \ell_2^-$ and $B_s \to \phi \ell_1^+ \ell_2^-$ a number of observables can be analysed.
The lepton forward-backward asymmetry $A_{FB}$, the $K^*$ polarization fractions, the angular distributions and the optimized $P$-observables \cite{Matias:2012xw} are particularly sensitive to NP effects.  For example, in  SM $A_{FB}$ has a zero  for  a quite precise value of $q^2$ \cite{Beneke:2000wa}, deviations would hint NP \cite{Ali:1991is,Ali:1994bf,Liu:1994cfa,Colangelo:1995jv,Ali:1999mm}. 
As for  LFV charged lepton decays, in the ABCD model   $\mu \to e \gamma$ arises from loops with $Z^\prime$, charged lepton couplings $\tau \to 3\mu$ and $\mu \to 3e$ occur at tree-level via $Z^\prime$ exchange, 
 $\mu \to e$ conversion in nuclei is mediated at tree-level by $Z^\prime$ and is sensitive to both lepton and quark couplings. The current experimental upper bounds on the branching fractions for such processes are presented in Table~\ref{tabLFV-leptons}.   $Z^\prime$   also provides a  contribution to   $(g-2)_{\mu,e}$.
Detailed discussions of the observables are in \cite{Aebischer:2019blw,Colangelo:2025nbd}.
\begin{table}[t]
\center{\begin{tabular}{|l|c|}
\hline
\quad {\rm LFV mode} &  ${\cal B}$ experiment \\
\hline
$\tau^- \to \mu^- \mu^+ \mu^-$ &  $<2.1 \times 10^{-8}$ \cite{Navas:PDG}\\
$\mu^- \to e^- \gamma$ &  $<1.5 \times 10^{-13}$ \cite{MEGnew}\\
$\mu^- \to e^- e^+ e^-$ & \,\,\,\, \,\,$<1.0 \times 10^{-12}$ \cite{SINDRUM:1987nra,Navas:PDG} \\
$\mu^- \to e^-$ conversion in $^{48}_{22}{\rm{Ti}}$ &  \quad \,\,  $<4.3 \times 10^{-12}$ \cite{Wintz:1998rp,Kaulard:1998rb}\\
\hline
\end{tabular}  }
\caption {\small  Experimental upper bounds (at $90 \, \%$ C.L.)  for  LFV charged lepton transitions.}
\label{tabLFV-leptons}
\end{table}

\section{Selected results}

We briefly describe the strategy  for the numerical analysis and the results  for selected observables   \cite{Colangelo:2025nbd}.
The  parameter space is at first constrained by experimental data on $\Delta F=2$ observables. The parameters  are those in  \eqref{pars} together with  the CKM and PMNS matrix elements. For the PMNS matrix  the results of  a global fit obtained assuming normal ordering for neutrinos are used \cite{Esteban:2024eli}. For the  four independent parameters of the CKM matrix, we choose $V_{us}$,  $|V_{cb}|$,  $|V_{ub}|$ and the phase $\gamma$. $V_{us}$ and $\gamma$  are  set to $V_{us}=0.2253$ and $\gamma=64.6^\circ$, 
  $|V_{cb}|$ and $|V_{ub}|$ are varied in the ranges spanned by the exclusive-inclusive determinations:
  \be
 |V_{cb}| \in \big[ |V_{cb}|_{exc},\,|V_{cb}|_{inc} \big] \,\, , \qquad |V_{ub}| \in \big[ |V_{ub}|_{exc},\,|V_{ub}|_{inc} \big] \,\,,
 \label{VcbVubRanges}
 \ee
 with  $|V_{cb}|_{exc}=(39.1 \pm 0.5)\times 10^{-3}$,  $|V_{cb}|_{inc}=(42.19 \pm 0.78)\times 10^{-3}$,  
  $|V_{ub}|_{exc}=(3.51 \pm 0.12)\times 10^{-3}$,  $|V_{ub}|_{inc}=(4.19 \pm 0.17)\times 10^{-3}$  \cite{HeavyFlavorAveragingGroupHFLAV:2024ctg}.
We vary $g_{Z^\prime} \in [0.01,\,0.1]$, and consider  $M_{Z^\prime}=1$ and $3$ TeV.

\begin{figure}[t]
\begin{center}
\includegraphics[width =0.3 \textwidth]{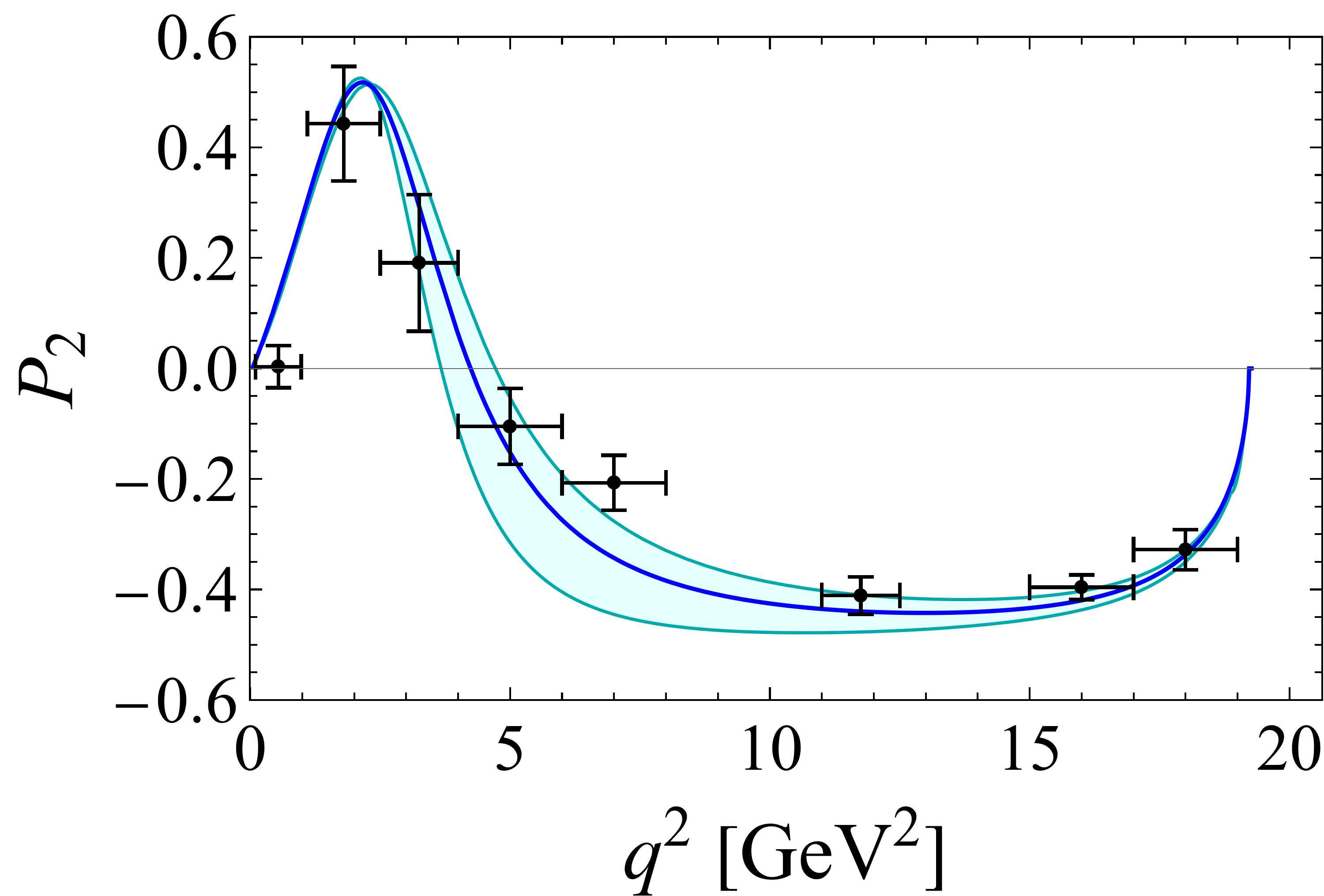}
\hskip 0.5cm \includegraphics[width =0.3 \textwidth]{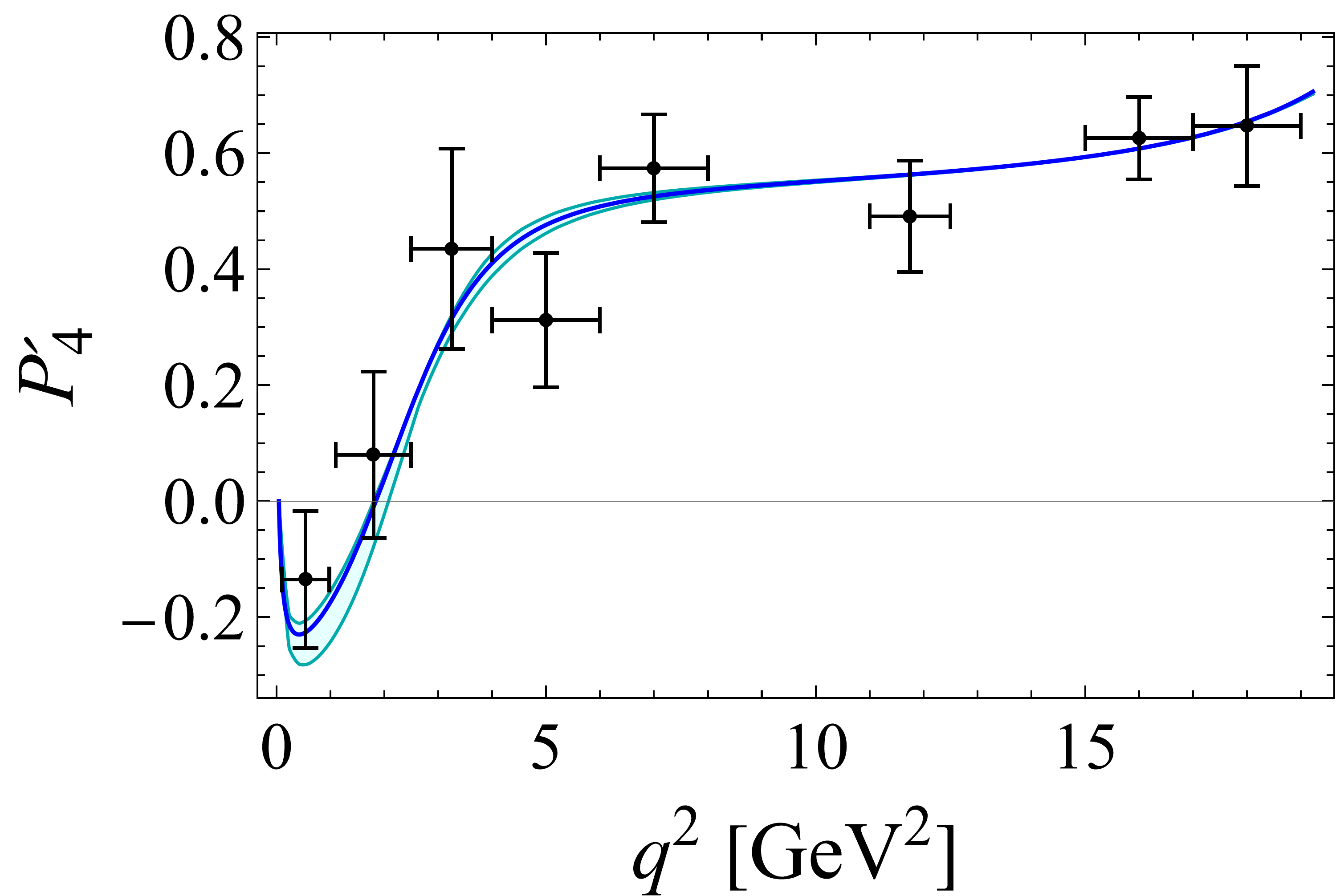}
\hskip 0.5cm \includegraphics[width =0.3 \textwidth]{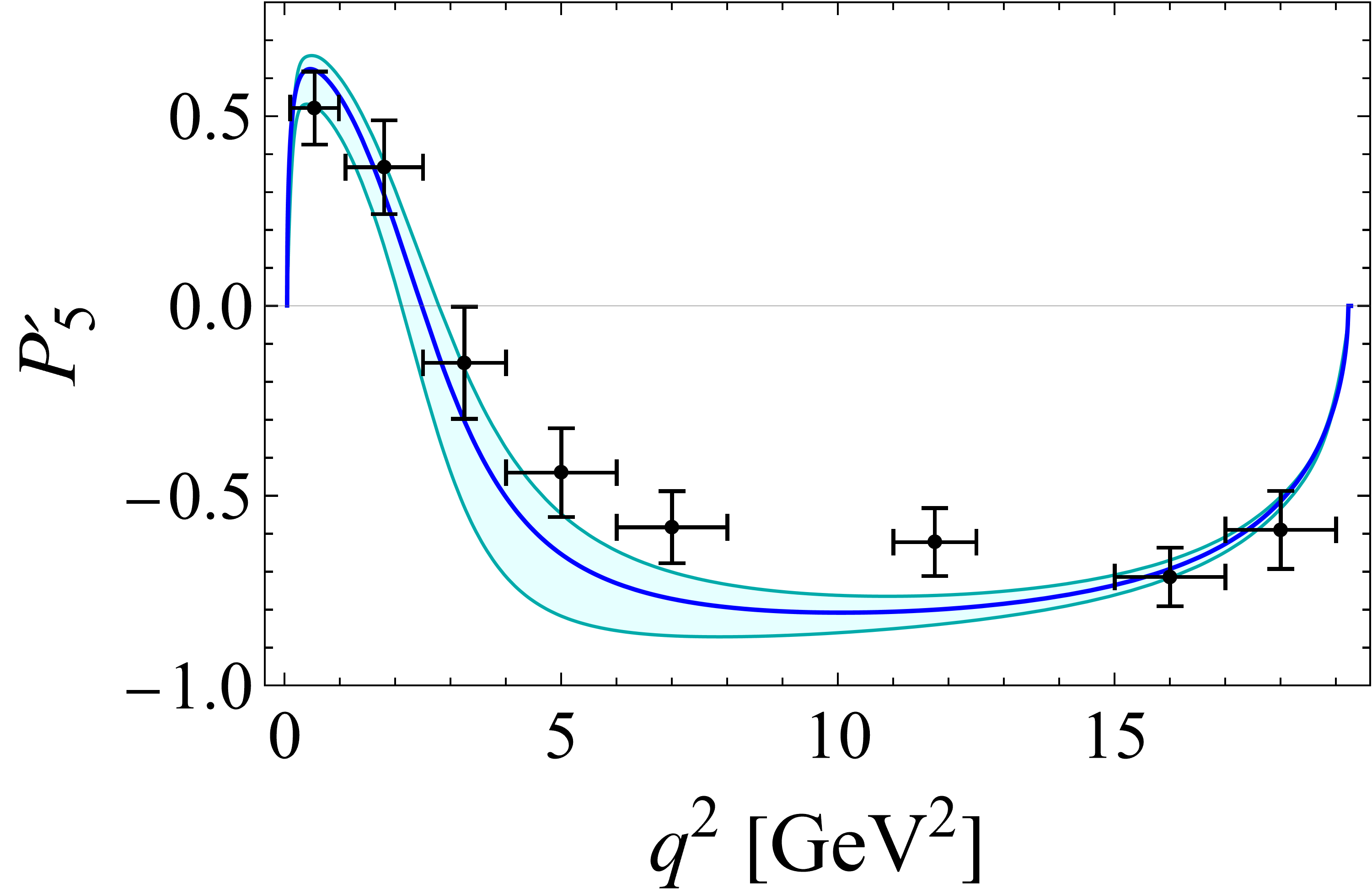}\\ 
\caption{\baselineskip 10pt  \small Angular $P$-observables for  the mode $\bar B^0 \to \bar K^{*0} (K \pi) \mu^+ \mu^-$. The black dots  correspond to LHCb measurement \cite{LHCb:2020lmf}. }\label{Pobs-mu}
\end{center}
\end{figure}

\begin{figure}[b]
\begin{center}
\includegraphics[width =0.4 \textwidth]{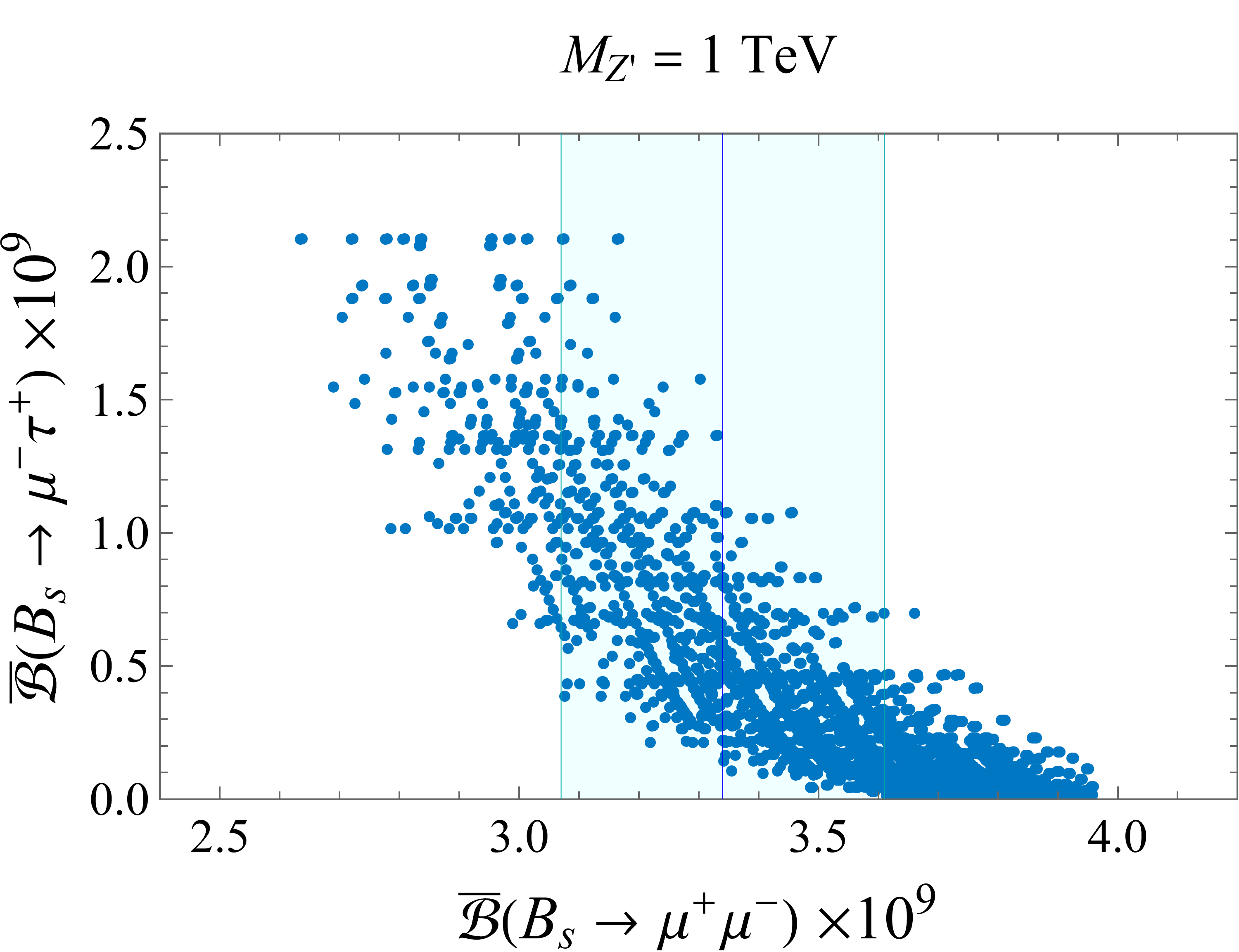} \hskip .4cm
\includegraphics[width =0.4 \textwidth]{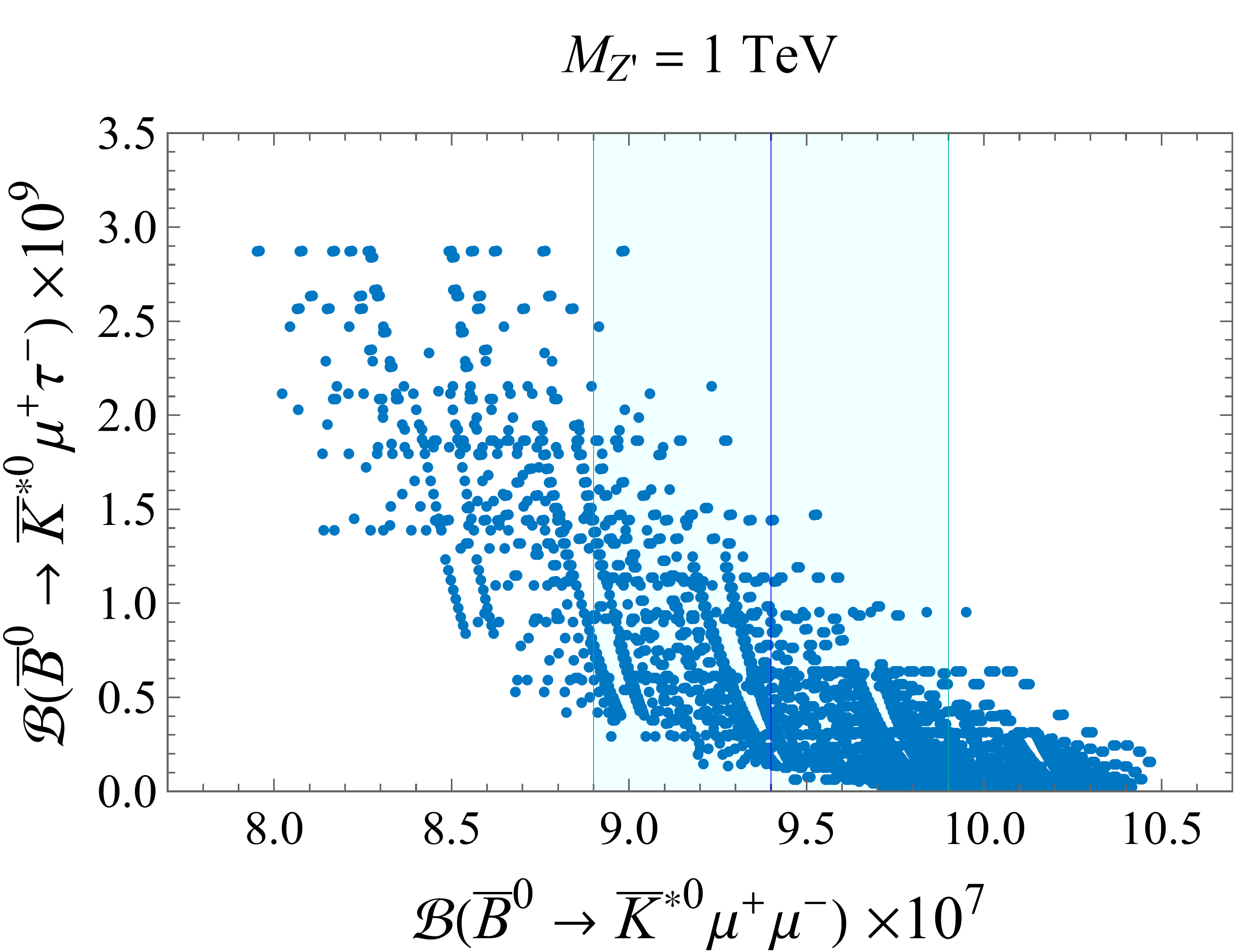} 
\caption{\baselineskip 10pt  \small  Correlation between the branching fractions ${\bar {\cal B}}(B_s \to \mu^+ \mu^-)$ and ${\bar {\cal B}}(B_s \to \tau^+ \mu^-)$  (left panel) and ${\cal B}(\bar B^0 \to \bar K^{*0} \mu^+ \mu^-)$ and ${\cal B}(\bar B^0 \to \bar K^{*0} \mu^+ \tau^-)$  (right panel) for $M_{Z^\prime}=1$ TeV, obtained after imposing $\Delta F=2$ constraints.  The shaded vertical band corresponds to the experimental world averages  \cite{Navas:PDG}. }\label{LFCvsLFV}
\end{center}
\end{figure}

The parameter space $(\epsilon_1, \epsilon_2)$ is constrained including  the $Z^\prime$  contribution to $\Delta F=2$ observables:
 the mass differences $\Delta M_d$, $\Delta M_s$, $\Delta M_K$ between the neutral mesons $B_d$, $B_s$ and kaons, the CP asymmetries $S_{\psi K_s}$, $S_{\psi \phi}$ and the CP violating parameter in the kaon sector $\epsilon_K$.
Such observables are related to the neutral meson  $M-{\bar M}$  mixing ($M = K^0,\,B_{d,s}$), which in  SM  occurs at
the one-loop level due to box diagrams mediated by internal up-type quark exchanges \cite{Buras:2020xsm}, while in the ABCD model  receive a tree-level contribution mediated by $Z^\prime$.
$\Delta M_d$, $\Delta M_s$ and the CP asymmetries $S_{J\psi K_s}$, $S_{\phi \psi}$ are required to stay  in a range around $5\%$ of the central value of the experimental result,   
$\Delta M_K$ is required to lie within 25$\%$ of the SM value (considering possible long-distance contributions),  $\epsilon_K$  in the range $\epsilon_K\in[2.0,\,2.5]\times 10^{-3}$ \cite{Navas:PDG}.
Such requirements determine the allowed  regions  in the  $(\epsilon_1,\,\epsilon_2)$ plane depending on   $M_{Z^\prime}$.
In the selected parameter space we at first investigate to which extent the lepton flavour conserving FCNC $B$ and $B_s$ decays are affected.
After imposing $\Delta F=2$ constraints, the NP contributions to the Wilson coefficients ${\rm Re}(C_9^{\rm NP})_{\ell \ell}$ and ${\rm Re}(C_{10}^{\rm NP})_{\ell \ell}$, $\ell=\mu,\,\tau$, are found to be at the  $\simeq 10\%$ level of the corresponding SM coefficients  \cite{Colangelo:2025nbd}. Deviations in LFC modes are found, but they are modest. As  stressed in \cite{Aebischer:2019blw}, in the ABCD model quark and lepton sectors mutually  preclude large deviations from  SM: this is consistent with the  small  deviations   from   SM  represented by the flavour anomalies.
As an example, we show in Fig.~\ref{Pobs-mu} the results for angular $P$-observables  in $\bar B^0 \to \bar K^{*0} (K \pi) \mu^+ \mu^-$, together with the measurements provided by the LHCb Collaboration \cite{LHCb:2020lmf}.
The form factors parametrizing  the $B \to K^*$ matrix elements of the operators in the Hamiltonian \eqref{hamil} are those determined in \cite{Bharucha:2015bzk};
the values of other input quantities  are quoted in \cite{Colangelo:2025nbd}.
We also study  the correlations between LFC and  the corresponding lepton-flavour-violating (LFV) modes. The LFC processes effectively bound  the LFV ones. 
In Fig.~\ref{LFCvsLFV} we display the  correlations between the branching fraction of  $B_s \to \ell_1^+ \ell_2^-$ and  $\bar B^0 \to \bar K^{*0} \ell_1^+ \ell_2^-$.

\begin{figure}[b]
\begin{center}
\includegraphics[width =0.47 \textwidth]{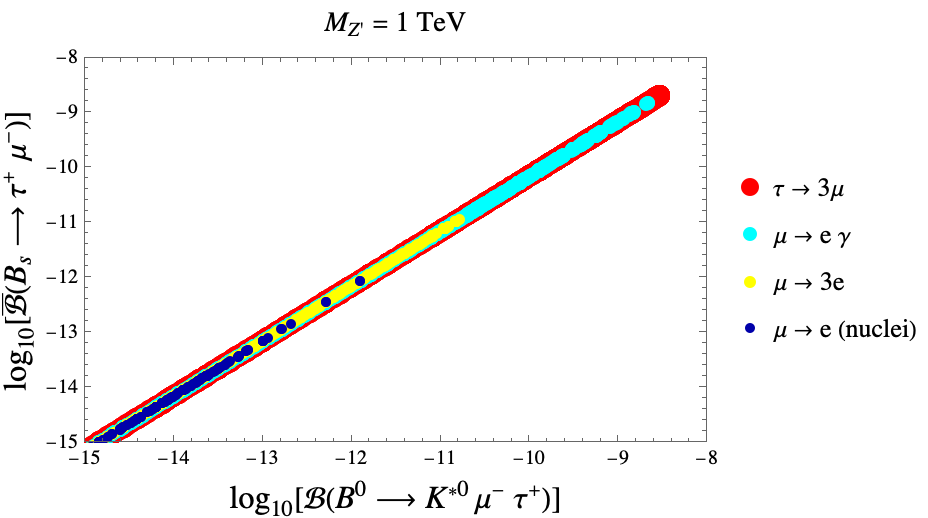} \hskip 0.4cm 
\includegraphics[width =0.47 \textwidth]{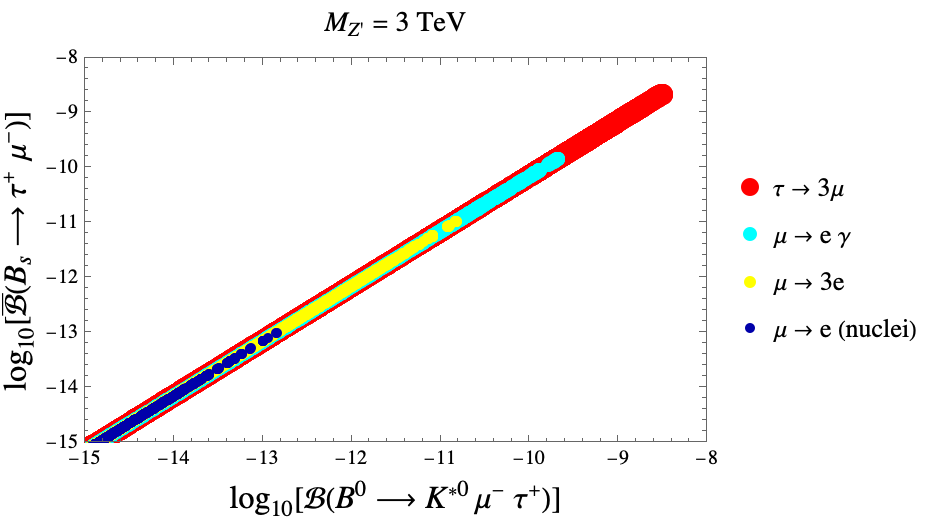} \\ 
\caption{\baselineskip 10pt  \small  Correlation between  ${\bar {\cal B}}(B_s \to \tau^+ \mu^-)$ and ${\cal B}(\bar B^0 \to \bar K^{*0} \tau^+ \mu^-)$ when the constraints from LFV charged lepton decay modes are subsequently imposed as specified in the legenda, for
$M_{Z^\prime}=1$ TeV (left panel) and $M_{Z^\prime}=3$ TeV (right panel).  
}\label{correlationfinal}
\end{center}
\end{figure}

Purely leptonic LFV processes can also  be  considered together with their correlations with LFV $B_{(s)}$ decays. Focusing on   $\tau^- \to \mu^- \mu^+ \mu^-$, $\mu^- \to e^- \gamma$, $\mu^- \to e^- e^+ e^-$ and $ \mu^- \to e^-$ conversion in nuclei, it is found that  the experimental upper bounds on such modes in Table \ref{tabLFV-leptons}  play a hierarchical role in constraining the branching fractions  of LFV $B$ and $B_s$ decays.  
 The bounds for $\tau^- \to \mu^- \mu^+ \mu^-$, $\mu^- \to e^- \gamma$  allow branching ratios for the LFV $B_{(s)}$ decays of ${\cal O}(10^{-9})$,  more stringent constraints stem from the experimental upper limits on   $\mu^- \to e^- e^+ e^-$ and $ \mu^- \to e^-$ conversion in nuclei,  pushing  the branching fractions down to  ${\cal O}(10^{-11})$ and ${\cal O}(10^{-13})$. This is summarized in Table~\ref{tabLFV1000} and in Fig.~\ref{correlationfinal}.
 %
 %%%%%%%%%%%%%%%
\begin{table}[t]
\center{\begin{tabular}{|l|c|c|c|c|c|}
\hline
\quad {\rm LFV mode} & ${\cal B}^{(1)}\times10^{9}$  & ${\cal B}^{(2)}\times10^{9}$ & ${\cal B}^{(3)}\times10^{11}$ & ${\cal B}^{(4)}\times10^{13}$ & experiment 
\\
\hline
$B_s \to \tau^+ \mu^-$ &  $0.00 \div  2.10$     &  $0.00 \div  1.60$ &    $0.00 \div  1.20$   &   $0.00 \div  9.20$    & $ <4.2 \times 10^{-5}$ \cite{Navas:PDG}\\
$\bar B^0 \to \bar K^{*0} \tau^+ \mu^-$ &  $0.00 \div 2.90$ &    $0.00 \div 1.15$ &    $0.00 \div  1.60$    & $0.00 \div  5.10$    & $<1.0 \times 10^{-5}$ \cite{Navas:PDG}
\\
$B_s \to \phi \tau^+ \mu^-$ &  $0.00 \div  3.43$ &      $0.00 \div  2.60$  &  $0.00 \div  1.95$& $0.00 \div  6.10$&  $<1.0 \times 10^{-5}$\cite{LHCb:2024wve} 
\\
\hline
$B_s \to \tau^+ \mu^-$ &   $0.00 \div 2.30$     & $0.00 \div 0.14$ &  $0.00 \div  1.10$  & $0.00 \div  1.05$ &$ <4.2 \times 10^{-5}$ \cite{Navas:PDG}
\\
$\bar B^0 \to \bar K^{*0} \tau^+ \mu^-$ &  $0.00 \div 3.10$ &   $0.00 \div 0.20$   &  $0.00 \div  1.53$   &  $0.00 \div  1.42$  &$<1.0 \times 10^{-5}$ \cite{Navas:PDG}
\\
$B_s \to \phi \tau^+ \mu^-$ &  $0.00 \div 3.70$ &     $0.00 \div 0.25$  & $0.00 \div  1.80$    &  $0.00 \div  1.70$   & $<1.0 \times 10^{-5}$\cite{LHCb:2024wve} 
\\
\hline
\end{tabular}  }
\caption {\small  Ranges of branching fractions of LFV $B$, $B_s$ decays in  the scenario A of the ABCD model. The first three rows correspond to  $M_{Z^\prime}=1$ TeV, the last three rows to $M_{Z^\prime}=3$ TeV. 
The four quoted ranges are obtained after imposing the constraints from the upper bounds on LFV charged  lepton decay modes  in Table~\ref{tabLFV-leptons}:
${\cal B}^{(1)}$    from $\tau^- \to \mu^- \mu^+ \mu^-$,  ${\cal B}^{(2)}$   from  ${\cal B}(\mu^- \to e^- \gamma)$,  ${\cal B}^{(3)}$    from  $\mu^- \to e^- e^+ e^-$,  ${\cal B}^{(4)}$    from  $\mu^- \to e^-$ conversion in titanium.   The last column includes the  experimental upper bounds   (at $90\%$ C.L.). }
\label{tabLFV1000}
\end{table}
%%%%%%%%%%%%%%%%%

\begin{figure}[t]
\begin{center}
\includegraphics[width =0.4 \textwidth]{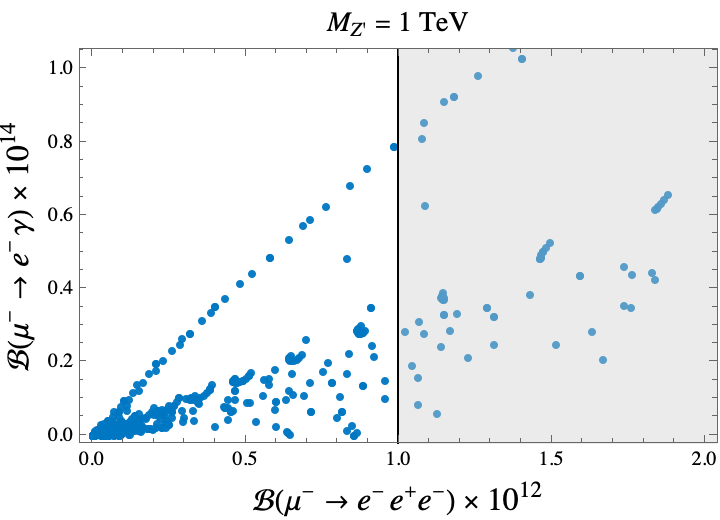} \hskip 0.4cm 
\includegraphics[width =0.4 \textwidth]{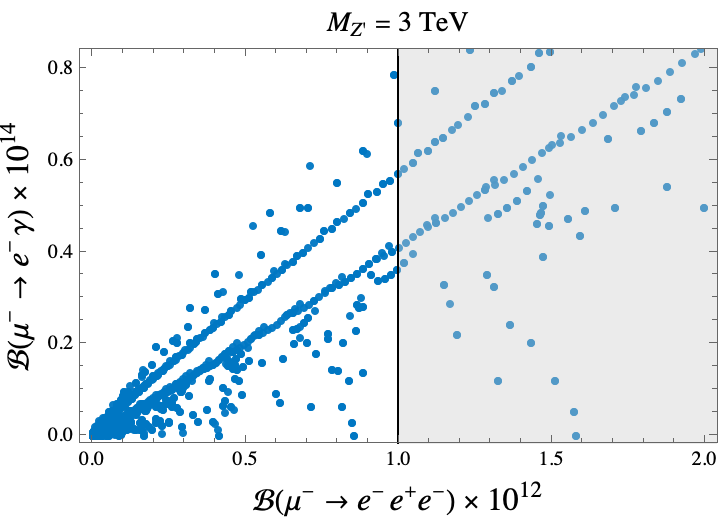} \\ \vskip 0.4cm 
\includegraphics[width = 0.4\textwidth]{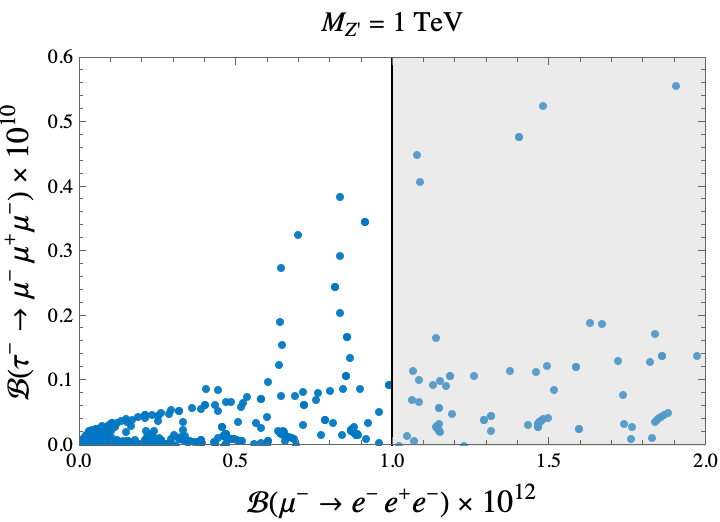} \hskip 0.4cm 
\includegraphics[width = 0.4\textwidth]{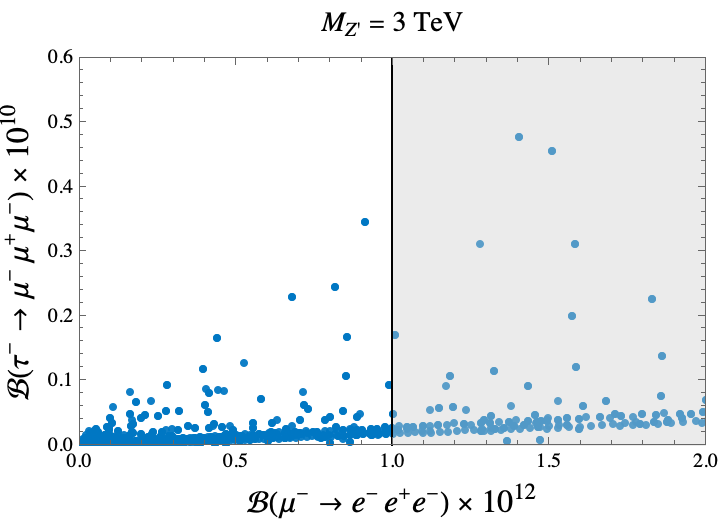}
\caption{\baselineskip 10pt  \small  Correlation between  ${\cal B}(\mu^- \to e^- e^+ e^-)$ and ${\cal B}(\mu^- \to e^- \gamma)$ (top panels) and 
 between ${\cal B}(\mu^- \to e^- e^+ e^-)$ and ${\cal B}(\tau^- \to \mu^- \mu^+ \mu^-)$  (bottom panels), for
$M_{Z^\prime}=1$ TeV (left panel) and $M_{Z^\prime}=3$ TeV (right panel).  The gray band corresponds to the upper bound ${\cal B}(\mu^- \to e^- e^+ e^-)<1 \times 10^{-12}\,\, (\rm{at}\,  90\% \, {\rm CL})$\cite{SINDRUM1988,Navas:PDG}.
}\label{leptondecaysvsmu3e}
\end{center}
\end{figure}

Among  other results, the path of correlations among  charged lepton flavour violating processes has been anaysed, an outcome is  shown in Fig.~\ref{leptondecaysvsmu3e}. The  size of the $Z^\prime$ contribution to $(g-2)_{\mu}$ turns out to be  at most comparable with the hadronic  light-by-light  contribution \cite{Colangelo:2025nbd}.

\section{Conclusions}
We have analyzed a simple extension of the SM,  the ABCD model,  which  predicts  a new gauge boson $Z^\prime$ with 
  generation-dependent   flavour conserving and flavour violating couplings to fermions \cite{Aebischer:2019blw}.  
  Requiring gauge anomaly cancellation, the quark and lepton couplings to $Z^\prime$ are related, so that  the observables in the two sectors are correlated.
Quark and lepton sectors are found to mutually act to exclude  large deviations from  SM,  a result  consistent with the small tensions observed in the  flavour anomalies.
On the other hand, SM forbidden processes, namely charged lepton flavour violating  transitions, can occur  at tree-level.
 Correlations arise, e.g.,  among leptonic LFV modes and  LFV decays of $B_{(s)}$ mesons \cite{Colangelo:2025nbd}.  Several signatures are predicted in this model, their tests  are part of the  physics programmes of current and planned experimental facilities.

\section*{Acknowledgements}
 F.D.F. thanks the coauthors of \cite{Aebischer:2019blw}. We thank A.J. Buras for very useful discussions.
This study has  been carried out within the INFN project (Iniziativa Specifica)  SPIF.

%\newpage
\bibliographystyle{JHEP}
\bibliography{refDFP}

\end{document}